# Multicanonical molecular dynamics by variable-temperature thermostats and variable-pressure barostats

Cheng Zhang[1, 2] and Michael W. Deem[1, 2, 3]

[1] Applied Physics Program,

[2] Department of Bioengineering, &

[3] Department of Physics

Rice University, Houston, TX 77005

Abstract

Sampling from flat energy or density distributions has proven useful in equilibrating complex systems with large energy barriers. Several thermostats and barostats are presented to sample these flat distributions by molecular dynamics. These methods use a variable temperature or pressure that is updated on the fly in the thermodynamic controller. These methods are illustrated on a Lennard-Jones system and a structure-based model of proteins.

# I. Introduction

Molecular dynamics (MD) is a useful tool to study complex molecular systems. A regular MD simply follows Newton's equation, which conserves the total energy $E$ and thus generates configurations in a microcanonical ensemble. We can, however, modify the equation of motion by an artificial thermostat[1-9] to sample a different distribution, in which the total energy fluctuates, such as the canonical distribution.

For a system with high energy barriers, the limited exploration of the energy landscape in the canonical distribution is insufficient. A multicanonical ensemble[10-13], in which the energy is broadly sampled, is more suitable, see Fig. **1**(a). The construction of the multicanonical ensemble is nontrivial, for the flat energy distribution requires the inverse of the unknown density of states, $1/\Omega(E)$, as the sampling weight.

We here pursue the construction of multicanonical[10-13] MD by using a variable temperature equal to $\beta(E) = d\log\Omega/dE$. We describe several thermostats to set this temperature. The method is built on a previous Monte Carlo (MC) method by Yan and de Pablo[14]. Here the sampling along the total energy is directly incorporated into the variable-temperature thermostat for the MD simulation.

In section II, we describe the method. In section III, we numerically test the method in a few examples. We conclude the article in section IV.

## II. Method

### II.A Microcanonical, canonical, and multicanonical ensembles

*Microcanonical ensemble*

In classical statistical mechanics, integration of Newton's equation generates configurations in a microcanonical ensemble under a fixed total energy $E = U(\mathbf{q}) + K(\mathbf{p})$, where $U(\mathbf{q})$ and $K(\mathbf{p}) = \mathbf{p}^2/2m$ are the potential and kinetic energies, respectively. Thus, configurations from an ergodic trajectory sum to the density of states:

$$\Omega(E) \propto \int_{\mathbf{q},\mathbf{p}} \delta[E - U(\mathbf{q}) - K(\mathbf{p})] d\mathbf{q}\, d\mathbf{p} \propto \int_{\mathbf{q}} [E - U(\mathbf{q})]^{N_f/2-1} \Theta[E - U(\mathbf{q})] d\mathbf{q}, \quad (1)$$

where $\Theta(x)$ is the step function, which is 1 if $x > 0$ or 0 otherwise, and $\delta(x) = \Theta'(x)$ is the $\delta$-function. We have also integrated out the $N_f$ momentum degrees of freedom[15]:

$$\int_{\mathbf{p}} \delta(E - U - \mathbf{p}^2/2m) d\mathbf{p} = \frac{\sqrt{2m\pi}^{N_f}}{\Gamma(N_f/2)} (E - U)^{N_f/2-1} \Theta(E - U),$$

with $\Gamma(x)$ being the gamma function. For a molecular system, $N_f$ is three times the number of particles less the number of conserved quantities, such as the total linear and angular momenta.

The microcanonical temperature, or the derivative of the entropy $S(E) = k_B \log \Omega(E)$ [16,17] with $k_B$ being the Boltzmann factor, can be evaluated as a configuration average[15,18]

$$\beta(E) \equiv \frac{d \log \Omega(E)}{dE} = \left\langle \frac{N_f/2 - 1}{K(\mathbf{p})} \right\rangle_E, \quad (2)$$

where $\beta(E) = 1/[k_B T(E)]$ is the inverse temperature, $\langle ... \rangle_E$ denotes an average under a fixed total energy $E$, and we have used Eq. (1):

$$\frac{1}{\Omega(E)}\frac{d\Omega(E)}{dE} = \int_{\mathbf{q}} \left(\frac{N_f/2-1}{E-U(\mathbf{q})}\right) \frac{[E-U(\mathbf{q})]^{N_f/2-1}\Theta[E-U(\mathbf{q})]}{\Omega(E)} d\mathbf{q} = \left\langle \frac{N_f/2-1}{K} \right\rangle_E .$$

Here we have assumed $N_f > 2$ so that the singularity at $E = U(\mathbf{q})$ is negligible. Other expressions for $\beta(E)$ also exist[18], as shown in Appendix A.

*Canonical ensemble*

The canonical ensemble is a superposition of microcanonical ensembles at various $E$, each with a weight $\exp(-\beta E)$. The overall $E$ distribution is given by

$$\rho(E) \propto \int_{\mathbf{q},\mathbf{p}} \delta[E - H(\mathbf{q},\mathbf{p})] d\mathbf{q}\, d\mathbf{p}\, \exp(-\beta E) = \Omega(E)\exp(-\beta E) . \tag{3}$$

To realize a canonical distribution in MD, one needs a thermostat[1-9] to alter the equation of motion either deterministically[3-8] or stochastically[1,2,8,9].

*Multicanonical ensemble*

We now consider a generalized ensemble composed of microcanonical ensembles with an arbitrary weight $\exp[-f(E)]$ instead of the Boltzmann weight $\exp(-\beta E)$. We approximate $f(E)$ by a piecewise linear function on a finely-spaced grid $E_0, E_1, \ldots, E_n$, as in Fig. **1**(b). Then within each $(E_k, E_{k+1})$, $f(E) \approx f(E_k) + f'(E_k)(E - E_k)$, and the ensemble can be treated as locally canonical with the effective temperature being $f'(E_k)$[10]. The constant $f(E_k) - f'(E_k)E_k$ does not distinguish configurations, and is therefore irrelevant. In the limit of infinitesimal intervals, the local effective temperature becomes $f'(E)$.

The multicanonical ensemble is the special case of the generalized ensemble described above, in which $f(E) = \log \Omega(E) + \text{const.}$, such that the $E$ distribution

$$\rho_{\text{muca.}}(E) = \int_{\mathbf{q},\mathbf{p}} \delta[E - H(\mathbf{q},\mathbf{p})] d\mathbf{q} \, d\mathbf{p} \, \exp[-f(E)] \propto \Omega(E) \exp[-\log \Omega(E)] = 1$$

is flat. Thus, the local temperature $f'(E)$ should be energy dependent, and it is equal to $(\log \Omega)'(E)$, which can be evaluated by Eq. (2).

Thus, molecular dynamics in the multicanonical ensemble can be achieved by a thermostat with an energy-dependent target temperature $f'(E) = (\log \Omega)'(E)$. Below, we describe how this can be done. As in the previous MC method[14], $\Omega(E)$ is estimated from its derivative by Eq. (2), instead of from the energy histogram[10,12,13]. Strictly speaking, detailed balance is violated if we constantly update $(\log \Omega)'(E)$ by Eq. (2) in simulation. Earl and Deem, however, showed that this practice, commonly used in the adaptive integration methods[14,19] and to be used here, satisfies balance in infinite-order Markov chains[20]. This updating practice converges in the present application, and so the present method satisfies balance at long times. Note also that the multicanonical ensemble here is formulated for the total energy[14], instead of the potential energy[10,12].

**II.B Algorithm**

We now describe an algorithm for a multicanonical MD. The aim is to sample a flat energy distribution in a given range $(E_{\min}, E_{\max})$. We split the range into $n$ intervals at $E_0 = E_{\min}$, $E_1$, ..., $E_n = E_{\max}$, such that each $(E_k, E_{k+1})$ accumulates data for computing the average of $(N_f/2 - 1)/K(\mathbf{p})$ to estimate $\beta(E)$ in Eq. (2):

$$\hat{\beta}(E) = \left\langle \frac{N_f/2 - 1}{K(\mathbf{p})} \right\rangle_{E_k < E < E_{k+1}}. \qquad (2')$$

The algorithm:

1. Integrate Newton's equation for $\Delta t_{MD}$.

2. Locate the interval $k$ that contains the current total energy $E$ (if $E_{min} < E < E_{max}$); and deposit $(N_f/2-1)/K(\mathbf{p})$ into the accumulator there.

3. Use Eq. (2′) to estimate $\hat{\beta}(E)$; but if the number of data points is less than 100, or if $E$ lies outside of $(E_{min}, E_{max})$, use a default value $\hat{\beta}(E) = \beta_d$, which should correspond to an energy $E \in (E_{min}, E_{max})$.

4. Use $\hat{\beta}(E)$ in one of the variable-temperature thermostats listed below.

5. Repeat Step 1 until the simulation ends.

*Remarks*

1. We show in Appendix A that the use of the interval-averaged formula Eq. (2′) to approximate $\beta(E)$ produces no systematic error: the estimated values of the density of states $\log \hat{\Omega}(E_k) = \int^{E_k} \hat{\beta}(E) dE$ at the boundaries $E_k$ asymptotically approach $\log \Omega(E_k)$ for any interval size $E_{k+1} - E_k$.

2. While the average in Eq. (2′) is invariant with respect to the interval size, the energy range, the number of intervals, and the number of data points will affect the rate of convergence. The number of intervals should be large enough so that the $\log \hat{\Omega}(E)$ is sufficiently close to $\log \Omega(E)$, i.e., the dotted line segments in

Fig. **1**(b) closely approximate the solid curve. Furthermore, the number of data points per interval should be sufficiently large to provide a reliable estimate of the average.

3. In Step 3, a constant $\hat{\beta}(E) = \beta_d$ is assumed for $E$ outside of $(E_{min}, E_{max})$ for simplicity. It, however, causes a large discontinuity there; and a better boundary condition may improve this. The number 100 is also arbitrary.

4. After simulation, we can recover the density of potential-energy state $g(U) = \int_\mathbf{q} \delta[U - U(\mathbf{q})] d\mathbf{q}$ from the observed distribution $h(U)$ of the potential energy $U$ by

$$g(U) = h(U) \bigg/ \left[ \int_U^{+\infty} (E-U)^{N_f/2-1} \big/ \hat{\Omega}(E) \, dE \right], \qquad (4)$$

see Appendix B for a proof. The denominator of Eq. (4) can be evaluated exactly by the incomplete gamma function[21], as $\log \hat{\Omega}(E)$ is piecewise linear. One can then compute the canonical partition function $Z(\beta) = \int g(U) \exp(-\beta U) dU$, and its derivatives. Eq. (4) also applies to a $\hat{\Omega}(E)$ that does not produce a flat $E$ distribution, see examples in Appendix A.

**II.C Variable-Temperature thermostats**

The thermostats for a canonical ensemble need to be modified as follows for a multicanonical MD. We list a few examples below. The dimension is denoted by $d$, which is usually 3.

*Nosé-Hoover chain thermostat*

The Nosé-Hoover chain equations[5,8] for a canonical ensemble are

$$\dot{\mathbf{q}} = \mathbf{p}/m, \quad \dot{\mathbf{p}} = \mathbf{F} - \zeta_1 \mathbf{p}, \quad \dot{\zeta}_1 = (\mathbf{p}^2/m - N_f/\beta)/Q_1 - \zeta_2\zeta_1,$$
$$\dot{\zeta}_k = (Q_{k-1}\zeta_{k-1}^2 - 1/\beta)/Q_k - \zeta_{k+1}\zeta_k, \quad (\text{for } k = 2,\ldots,M-1) \quad (5)$$
$$\dot{\zeta}_M = (Q_{M-1}\zeta_{M-1}^2 - 1/\beta)/Q_M,$$

where $\zeta_k$ are the thermostat variables, and $Q_k$ their masses. The corresponding Liouville's equation

$$\partial\rho/\partial t + \partial(\dot{\mathbf{q}}\rho)/\partial\mathbf{q} + \partial(\dot{\mathbf{p}}\rho)/\partial\mathbf{p} + \sum_{k=1}^{M}\partial(\dot{\zeta}_k\rho)/\partial\zeta_k = 0 \quad (6)$$

has a stationary ($\partial\rho/\partial t = 0$) solution

$$\rho(\mathbf{q},\mathbf{p},\{\zeta_k\}) \propto \exp[-\beta U(\mathbf{q}) - \beta\mathbf{p}^2/2m - \beta\sum_{k=1}^{M}Q_k\zeta_k^2/2],$$

which is indeed canonical[5].

For a multicanonical ensemble, we change the equation for $\dot{\zeta}_1$ in Eqs. (5) to

$$\dot{\zeta}_1 = \{[\hat{\beta}(E)/\beta_0]\mathbf{p}^2/m - N_f/\beta_0\}/Q_1 - \zeta_2\zeta_1, \quad (5')$$

and we replace the $\beta$ in the equations for $\dot{\zeta}_k$, $k \geq 2$, by $\beta_0$, such that the new stationary solution of Eq. (6) is

$$\rho(\mathbf{q},\mathbf{p},\{\xi_k\}) \propto \exp\{-\log\hat{\Omega}[U(\mathbf{q}) + \mathbf{p}^2/2m] - \beta_0\sum_{k=1}^{M}Q_k\zeta_k^2/2\}, \quad (7)$$

where $\beta_0$ is a reference inverse temperature, against which the set point of the system's inverse temperature is scaled in the term in braces in Eq. (5'). That is, with the use of $\hat{\beta}(E)$ in Eq. (5'), the stationary distribution is the desired Eq. (7), no matter the value of $\beta_0$. One expects that $\beta_0$ values near the average of $\hat{\beta}(E)$ will be efficient for sampling.

We recover the original Nosé-Hoover thermostat by setting $\zeta_2 = 0$, and removing the equations for $\dot\zeta_k$ with $k \geq 2$ and for $\dot\zeta_M$.

*Velocity-Rescaling thermostat*

The velocity-rescaling thermostat[9] achieves a canonical distribution by re-sampling the distribution of the total energy $E$ under fixed **q** coordinates:

$$\rho(E)\,dE = [E - U(\mathbf{q})]^{N_f/2-1} \exp(-\beta E)\,dE,$$

cf. Eqs. (1) and (3). The change $E \to E'$, hence $K \to K'$, is realized by scaling the velocity or momentum: $\mathbf{p} \to \sqrt{K'/K}\,\mathbf{p}$, for $U(\mathbf{q})$ is fixed. The kinetic energy is randomly updated and follows the Langevin equation[9]

$$dK/dt = \tfrac{1}{2}N_f/\beta - K + \sqrt{2K/\beta}\,\xi, \tag{8}$$

where $\xi$ is a Gaussian white noise that satisfies $\langle\xi(t)\xi(t')\rangle = \delta(t-t')$.

For a multicanonical ensemble, we modify Eq. (8) to

$$dK/dt = \tfrac{1}{2}N_f/\beta_0 - [\hat\beta(E)/\beta_0]K + \sqrt{2K/\beta_0}\,\xi, \tag{8'}$$

such that the corresponding Fokker-Planck equation

$$\partial\rho/\partial t + \partial(\{\tfrac{1}{2}N_f - [\hat\beta(E)/\beta_0]K\}\rho)/\partial K - \partial^2(K\rho/\beta_0)/\partial K^2 = 0$$

has the desired stationary solution $\rho(K) \propto K^{N_f/2-1}/\hat\Omega(U+K)$. Here, $\beta_0$ is a reference inverse temperature.

Note that Eq. (8) or (8′) can be integrated with a different time step from the MD one.

*Monte-Carlo velocity-rescaling thermostat*

We can also replace Eq. (8′) by an explicit MC move in the $E$, or $K$ space[14]. We first propose a change of $\log K$ by $\delta \in (-\Delta, \Delta)$, then accept the new $K' = Ke^{\delta}$ by

$$A(K \to K') = \min\left\{1, \frac{\hat{\Omega}(K+U)}{\hat{\Omega}(K'+U)}\left(\frac{K'}{K}\right)^{N_f/2}\right\}.$$

If $K'$ is accepted, it is also realized as a velocity rescaling: $\mathbf{p} \to \sqrt{K'/K}\,\mathbf{p}$.

*Langevin dynamics*

Langevin dynamics[1,17] modifies Newton's equation of motion by introducing a velocity damping term and a thermal noise:

$$\dot{\mathbf{q}} = \mathbf{p}/m, \quad \dot{\mathbf{p}} = \mathbf{F} - \zeta\mathbf{p} + \sqrt{2m\zeta/\beta}\,\boldsymbol{\xi}, \tag{9}$$

where $\zeta$ is a damping constant, and $\boldsymbol{\xi}$ is a Gaussian white noise satisfying $\langle \xi_i(t)\xi_j(t')\rangle = \delta(t-t')\delta_{ij}$.

For a multicanonical ensemble, we change Eq. (9) to

$$\dot{\mathbf{q}} = \mathbf{p}/m, \quad \dot{\mathbf{p}} = \mathbf{F} - [\hat{\beta}(E)/\beta_0]\zeta\mathbf{p} + \sqrt{2m\zeta/\beta_0}\,\boldsymbol{\xi}, \tag{9′}$$

such that the corresponding Fokker-Planck equation

$$\partial\rho/\partial t + \partial[(\mathbf{p}/m)\rho]/\partial\mathbf{q} + \partial\{[\mathbf{F} - \hat{\beta}(E)/\beta_0\,\zeta\mathbf{p}]\rho\}/\partial\mathbf{p} - \partial^2[(m\zeta/\beta_0)\rho]/\partial\mathbf{p}^2 = 0$$

has the desired stationary ($\partial\rho/\partial t = 0$) solution $\rho(\mathbf{q},\mathbf{p}) \propto 1/\hat{\Omega}[U(\mathbf{q}) + \mathbf{p}^2/2m]$. Here, $\beta_0$ is a reference inverse temperature.

*Andersen thermostat*

In the Andersen thermostat[2,8], we replace the velocity or momentum $\mathbf{p}_i^{old}$ of a random particle *i* by a Gaussian random vector drawn from the Maxwell distribution:

$$p_{\text{Maxwell}}(\beta; \mathbf{p}_i^{new}) = (\beta/2\pi m)^{d/2} \exp[-\beta (\mathbf{p}_i^{new})^2/2m]. \tag{10}$$

For a multicanonical ensemble, we treat $\mathbf{p}_i^{new}$ from Eq. (10), with $\beta$ replaced by $\hat{\beta}(E)$, as a MC trial, and accept it by the Metropolis probability

$$A(\mathbf{p}_i^{old} \to \mathbf{p}_i^{new}) = \min\left\{1, \; \frac{\hat{\Omega}(E)}{\hat{\Omega}(E')} \frac{p_{\text{Maxwell}}(\hat{\beta}(E'); \mathbf{p}_i^{old})}{p_{\text{Maxwell}}(\hat{\beta}(E); \mathbf{p}_i^{new})}\right\}, \tag{11}$$

where $E' = E + (\mathbf{p}_i^{new})^2/2m - (\mathbf{p}_i^{old})^2/2m$, and the two Maxwell factors $p_{\text{Maxwell}}(\ldots)$, defined in Eq. (10), offset the *a priori* sampling bias. For the canonical ensemble, $\hat{\beta}(E) = \hat{\beta}(E') = \beta$, $\hat{\Omega}(E)/\hat{\Omega}(E') = \exp[\beta(\mathbf{p}_i^{old})^2/2m - \beta(\mathbf{p}_i^{new})^2/2m]$, and Eq. (11) yields unity so that the trial is always accepted.

**II.D Volume distribution**

We can extend the above sampling strategy to explore a broad volume distribution, which is useful in studying a liquid-gas phase transition[22,23]. A flat volume distribution can be inefficient, however, as drastic free energy changes occur only at a high-density, or small-volume, regime. We therefore sample the distribution $\rho(V) \propto 1/V^\alpha$. To sample a flat density distribution, $\rho_n(n) = \text{const.}$, with $n = N/V$, we use $\alpha = 2$, because

$$\rho(V)\,dV = \rho_n(n)|dn| = \rho_n(n)\,N\,dV/V^2 \propto (1/V^2)\,dV.$$

To sample the distribution $\rho(V) \propto V^{-\alpha}$, we replace the constant target pressure in an isothermal-isobaric simulation[2,5,7,8,24-26] by the variable

$$\hat{p}(V) = \frac{\langle p_{\text{int.}}(\mathbf{q},\mathbf{p}) V^{\alpha} \rangle_{V_k < V < V_{k+1}}}{\langle V^{\alpha} \rangle_{V_k < V < V_{k+1}}}, \qquad (12)$$

where $p_{\text{int.}}(\mathbf{q},\mathbf{p}) = (\mathbf{p}^2/m + \mathbf{F} \cdot \mathbf{q})/(d \cdot V) - d\phi_{\text{tail}}(V)/dV$, where $\phi_{\text{tail}}(V)$ is the energy tail correction, such that the potential energy $U(\mathbf{q},V) = U_{\text{trunc.}}(\mathbf{q}) + \phi_{\text{tail}}(V)$[7,8,24-26]. The weight $V^{\alpha}$ in Eq. (12) ensures unbiased free energy $\hat{F}(V_k) = -\int_{V_{\min}}^{V_k} \hat{p}(V) dV$ for a finite interval width $V_{k+1} - V_k$, see Appendix A.

*Nosé-Hoover chain barostat*

To sample $\rho(V) \propto V^{-\alpha}$, we modify the Nosé-Hoover chain[5,7,8] equation for an isothermal-isobaric ensemble to

$$\begin{aligned}
\dot{\mathbf{q}} &= \mathbf{p}/m + \eta \mathbf{q}, \quad \dot{\mathbf{p}} = \mathbf{F} - (\zeta + \eta)\mathbf{p}, \quad \dot{V} = d\eta V, \\
\dot{\eta} &= \{[p_{\text{int.}}(\mathbf{q},\mathbf{p}) - \hat{p}(V)]V + (1-\alpha)/\beta\}d/W - \zeta_1 \eta, \\
\dot{\zeta_1} &= [\mathbf{p}^2/m + W\eta^2 - (N_f + 1)/\beta]/Q_1 - \zeta_2 \zeta_1, \\
\dot{\zeta_k} &= (Q_{k-1}\zeta_{k-1}^2 - 1/\beta)/Q_k - \zeta_{k+1}\zeta_k, \quad (\text{for } k = 2,\ldots,M-1) \\
\dot{\zeta_M} &= (Q_{M-1}\zeta_{M-1}^2 - 1/\beta)/Q_M,
\end{aligned} \qquad (13)$$

where $\eta$ and $\zeta_k$ are the variables for the barostat and thermostat, respectively, with $W$ and $Q_k$ being their respective masses. The corresponding Liouville's equation

$$\partial \rho/\partial t + \partial(\dot{\mathbf{q}}\rho)/\partial \mathbf{q} + \partial(\dot{\mathbf{p}}\rho)/\partial \mathbf{p} + \partial(\dot{\eta}\rho)/\partial \eta + \sum_{k=1}^{M} \partial(\dot{\zeta_k}\rho)/\partial \zeta_k + \partial(\dot{V}\rho)/\partial V = 0$$

has the desired stationary solution:

$$\rho(\mathbf{q},\mathbf{p},\eta,\{\zeta_k\},V) \propto \exp\{-\beta[U(\mathbf{q},V) + \mathbf{p}^2/2m + W\eta^2/2 + \sum_{k=1}^{M} Q_k \zeta_k^2/2 - \hat{F}(V)]\}V^{-\alpha},$$

where $\hat{F}(V) = -\int^V \hat{p}(V')dV'$ is the estimated Helmholtz free energy. The term $1-\alpha$ in the equation for $\dot{\eta}$ accounts for the ensemble weight $V^{-\alpha}$. We recover an isothermal-isobaric sampling by setting $\hat{p}(V)$ to a constant and $\alpha \to 0$. The $1-\alpha \to 1$ corrects the small error in the original Nosé-Hoover barostat[5,7,8].

We recover the Nosé-Hoover case by setting $\zeta_2 = 0$, and removing the equations for $\dot{\zeta}_k$ with $k \geq 2$ and for $\dot{\zeta}_M$.

### Langevin barostat

Alternatively, the pressure can be controlled by a Langevin equation[26] separate from the thermostat. The Langevin barostat is a stochastic version of Eq. (13):

$$\dot{\mathbf{q}} = \mathbf{p}/m + \eta \mathbf{q}, \quad \dot{\mathbf{p}} = \mathbf{F} - \eta \mathbf{p}, \quad \dot{V} = d\eta V, \quad (14)$$
$$\dot{\eta} = \{[p_{int.}(\mathbf{q},\mathbf{p}) - \hat{p}(V)]V + (1-\alpha)/\beta\}d/W - \zeta\eta + \sqrt{2\zeta/(\beta W)}\,\xi.$$

where $\xi$ is a Gaussian white noise satisfying $\langle \xi(t)\xi(t')\rangle = \delta(t-t')$, $\zeta$ and $W$ are two constants. The Fokker-Planck equation

$$\partial\rho/\partial t + \partial(\dot{\mathbf{q}}\rho)/\partial\mathbf{q} + \partial(\dot{\mathbf{p}}\rho)/\partial\mathbf{p} + \partial(\dot{V}\rho)/\partial V + \partial[(\dot{\eta})_{det.}\rho]/\partial\eta - \partial^2[(\zeta/\beta W)\rho]/\partial\eta^2 = 0,$$

where $(\dot{\eta})_{det.} = \{[p_{int.}(\mathbf{q},\mathbf{p}) - \hat{p}(V)]V + (1-\alpha)/\beta\}d/W - \zeta\eta$, admits the desired stationary distribution:

$$\rho(\mathbf{q},\mathbf{p},\eta,V) \propto \exp\{-\beta[U(\mathbf{q},V) + \mathbf{p}^2/2m + W\eta^2/2 - \hat{F}(V)]\}V^{-\alpha}.$$

### Explicit Monte-Carlo sampling

The above Langevin equation can be replaced by a MC move[8,22]: we first propose a change of $\log V^8$ by $\delta \in (-\Delta, \Delta)$, then accept the new $V' = Ve^\delta$ by

$$A(V \to V') = \min\left\{1, \frac{\exp\{-\beta[U(\sqrt[d]{V'/V}\mathbf{q},V') + (\sqrt[d]{V/V'}\mathbf{p})^2/2m - \hat{F}(V')]\}V'^{-\alpha}}{\exp\{-\beta[U(\mathbf{q},V) + \mathbf{p}^2/2m - \hat{F}(V)]\}V^{-\alpha}}\right\}. \quad (15)$$

If the move is accepted, we change the coordinates as $\mathbf{q} \to \sqrt[d]{V'/V}\,\mathbf{q}$ [8], and momenta as $\mathbf{p} \to \sqrt[d]{V/V'}\,\mathbf{p}$. The phase-space volume element $d\mathbf{q}\,d\mathbf{p}$ is conserved.

## III. Examples

### III.A Flat energy distribution

We tested the method in the $N = 108$ particle Lennard-Jones system[8]. The potential was cut off at $r_c = 2.5$ Å and shifted to zero. The density was $n = N/V = 0.7$. Each trajectory was integrated for $5\times 10^8$ steps; and the time step was $\Delta t_{MD} = 0.001$ unless specified otherwise. The energy $E$ range was $(-4N, 2N)$, and the interval was $0.02N$. The default inverse temperature $\beta_d$ for $E$ outside of the range or insufficient data and reference inverse temperature $\beta_0$ were both 1.0.

The energy distributions $\rho(E)$ from simulations using several variable-temperature thermostats are shown in Figs. **2**(a)-(e). The flatness of the energy distributions $\rho(E)$ helped detect if the variable-temperature thermostat was correct. The modifications in Sec. II.C were necessary to produce flat distributions. That is, if the $\beta$ was simply replaced by $\hat{\beta}(E)$ in Eqs. (5), (8), (9), and (10), flat energy distributions were not achieved, as shown in Fig. **2**(a), **2**(c), **2**(e), and **2**(f), respectively.

The thermostat equations are stiff if the mass $Q_1$ in the Nosé-Hoover chain is not extensive. An efficient $Q_1$ should be proportional to the number $N_f$ of degrees of freedom[4,6,25], which can be seen from the scales of the equation for $\dot{\zeta}_1$ in Eq. (5′). From the left side, we have $\rho(\zeta_1) \propto \exp(-\beta_0 Q_1 \zeta_1^2/2)$ by Eq. (7). Thus, $\zeta_1 \propto 1/\sqrt{Q_1}$ and $\dot{\zeta}_1 \sim \zeta_1/\Delta t_{MD} \sim 1/(\Delta t_{MD}\sqrt{Q_1})$. From the right side of Eq. (5′),

$[\hat{\beta}(E)/\beta_0]\mathbf{p}^2/m - N_f/\beta_0 \sim \sqrt{N_f}$, since it is proportional to the standard deviation of $\mathbf{p}^2/m$. Thus, the right side $\sim \sqrt{N_f}/Q_1$. Equating the two results, we find $\sqrt{Q_1}/\Delta t_{MD} \sim \sqrt{N_f}$. The other $Q_k$, $k \geq 2$, should be of order unity[6,25]. With $Q_1 = 10$, which was much smaller than $N_f$, visible deviations from the flat distribution can be seen, but a smaller MD time step $\Delta t_{MD}$ recovered a flat distribution, see Fig. **2**(f), as it must by Eq. (7).

The approximate[15] temperature formula $\tilde{\beta}(E) = N_f/\langle 2K \rangle_E$ [14,27] did not produce a flat energy distribution, but rather a $\rho(E) \propto 1/\langle K \rangle_E$, as shown in Fig. **2**(d). This is because $\tilde{\beta}(E)$ corresponds to a modified density of states weighted by $2K$, as discussed in Appendix A.

### III.B Structure-Based protein model

As an application of multicanonical MD, we studied the protein villin headpiece, Protein DataBank code: 1VII[28,29], by a simplified structure-based model[30-32]. Details of the model are described in Appendix C. Each trajectory was run for $2 \times 10^9$ steps with $\Delta t_{MD} = 0.002$. We used the MC velocity-rescaling variable-temperature thermostat with $\Delta = 0.05$. We show below that the thermodynamics of the model depends critically on the cutoff distance $r_c$ that defines contacts; the interaction is attractive between contact atoms, but repulsive otherwise.

As shown in Fig. **3**(a), the microcanonical temperature functions $\beta(E)$ under several different $r_c$ were similar to pressure-volume isotherms around a liquid-gas transition in a simple fluid[17,33]. The back-bending $S$-loop[34] that signifies a first-order phase transition occurred only when $r_c \geq 7$ Å. The sensitivity to $r_c$ may explain that both the presence and absence of $S$-loop were previously observed in similar models[13,35]. A strong first-order transition in the energy space may be absent with the common definitions of contacts[31,32,36], whose equivalent $r_c$ were about 4 to 5Å.

For $r_c > 6.5$ Å, the energy distributions $\rho(E)$ around the transition temperature $\beta_E^*$ were bimodal. Here, $\beta_E^*$ was defined as the temperature at which the two modes share the same height. See Fig. **3**(c) for the case of $r_c = 8$ Å. The free energy profile $-\log \hat{\Omega}(E) + \beta_E^* E = \int^E [\beta_E^* - \hat{\beta}(E')] dE'$ indeed had two basins.

We computed the density of potential-energy state $g(U)$ by the reweighting formula Eq. (4). The transition temperature $\beta_U^*$ determined from $g(U)$ was similar to $\beta_E^*$, see Fig. **3**(d). The free energy profile $-\log g(U) + \beta_U^* U$ along $U$ showed a higher barrier than that along $-\log \hat{\Omega}(E) + \beta_E^* E$.

**III.C Flat density distribution**

We tested the extension in Sec. II.D for the flat density distribution also on the 108 particle Lennard-Jones system. The pair potential was cut off at half-box length to facilitate volume change moves[8]. The density range was (0.05, 0.75) with a spacing 0.002. The default pressure $p_d$ for a density outside of the range or insufficient data was 0.1. $T = 1.2$. Each trajectory was run for $4 \times 10^8$ steps with $\Delta t_{MD} = 0.001$.

The density distribution, pressure-density curves, and free energy profile are shown in Fig. **4** for the Nosé-Hoover chain Eq. (13), Langevin Eq. (14), and MC Eq. (15) barostats. The density distributions were flat. The obtained transition pressures were 0.0696, 0.0695 and 0.0697 for the Nosé-Hoover chain, Langevin and MC barostats, respectively. The free energy profiles obtained at the transition pressures also agreed with one another, see Fig. **4**(b).

## IV. Discussion and Conclusion

A multicanonical, or flat-energy-distribution, molecular dynamics can be conveniently realized by a variable-temperature thermostat. Similarly, a flat-density-distribution MD can be realized by a variable-pressure barostat. The method is easy to implement with minimal modifications from a regular constant temperature or volume simulation. We have demonstrated the method of a Lennard-Jones system and a structure-based model of the villin headpiece.

The method computes the density of states $\Omega(E)$ [14] instead of the density of potential-energy states $g(U)$ [10,12]. The reweighting formula Eq. (4) is used to recover $g(U)$. In this way, thermal averages in the canonical ensemble are computed after simulation.

The method is somewhat more robust, in terms of achieving a flat energy distribution, with a smaller energy interval $\Delta E = E_{k+1} - E_k$, although it is formally correct for any $\Delta E$, as shown in Appendix A. A smaller interval also yields a finer $\hat{\Omega}(E_k)$, but may slow down the on the fly construction of the ensemble weight. The integral-identity technique[29,37], which allows borrowing data from neighboring intervals, and the adaptive averaging technique from the continuous tempering[29] may help improve convergence.

The multicanonical ensemble is commonly used in Monte Carlo and molecular dynamics simulations to overcome energy barriers. We here have described a number of variable-temperature thermostats and variable-pressure barostats to achieve flat energy or density distributions. We expect these methods may prove useful in simulating systems with complex energy landscapes.

## Acknowledgements

This work was supported by the Shared University Grid at Rice funded by NSF under Grant EIA-0216467 and by the US Department of Energy grant number DE-FG02-03ER15456.

## Appendix A. Unbiased density of states at interval boundaries

We show here that the values of the density of states $\hat{\Omega}(E_k) = \int^{E_k} \hat{\beta}(E)\,dE$ at the boundaries points $E_k$ are unbiased, although $\hat{\beta}(E)$ is an approximate average of $\beta(E)$ over finite intervals $(E_k, E_{k+1})$. This is because[29]

$$\frac{\Omega(E_{k+1})}{\hat{\Omega}(E_{k+1})} - \frac{\Omega(E_k)}{\hat{\Omega}(E_k)} = \int_{E_k}^{E_{k+1}} \frac{d}{dE}\left[\frac{\Omega(E)}{\hat{\Omega}(E)}\right]dE$$

$$= \int_{E_k}^{E_{k+1}} [(\log\Omega)'(E) - (\log\hat{\Omega})'(E)]\frac{\Omega(E)}{\hat{\Omega}(E)}dE$$

$$= h_k\left\langle\frac{N_f/2-1}{K(\mathbf{p})} - \hat{\beta}(E)\right\rangle_{E_k<E<E_{k+1}},$$

where we have used $(\log\Omega)'(E) = \langle(N_f/2-1)/K(\mathbf{p})\rangle_E$, $(\log\hat{\Omega})'(E) = \hat{\beta}(E)$. The ratio $\Omega(E)/\hat{\Omega}(E) = \int\delta[E-H(\mathbf{q},\mathbf{p})]d\mathbf{q}\,d\mathbf{p}/\hat{\Omega}(E)$ is the distribution density of the overall $E$ distribution, and we have defined $h_k \equiv \int_{E_k}^{E_{k+1}} \Omega(E)/\hat{\Omega}(E)\,dE$. Since the right hand side is zero by Eq. (2'), the left side vanishes. So $\hat{\Omega}(E_{k+1})/\hat{\Omega}(E_k) = \Omega(E_{k+1})/\Omega(E_k)$, and by recursion, $\hat{\Omega}(E_k)/\hat{\Omega}(E_0) = \Omega(E_k)/\Omega(E_0)$. Thus, $\hat{\Omega}(E_k)$ differs from $\Omega(E_k)$ only by a multiple.

The alternative formulas for $\hat{\beta}(E)$: $\tilde{\beta}^{(\mathbf{P})}(E) = N_f/\langle 2K(\mathbf{p})\rangle_{E_k<E<E_{k+1}}$ and $\tilde{\beta}^{(\mathbf{F})}(E) = \langle\nabla^2 U\rangle_{E_k<E<E_{k+1}}/\langle\mathbf{F}\cdot\mathbf{F}\rangle_{E_k<E<E_{k+1}}$ in ref. 14,27, have a similar property, although they do not produce flat $E$ distributions, even with infinitely small intervals. To see this, consider an arbitrary vector field $\mathbf{v} = \mathbf{v}(\mathbf{p},\mathbf{q})$. We define

$$\Omega_D(E) \equiv \int(\mathbf{v}\cdot\nabla H)\,\delta[E-H(\mathbf{q},\mathbf{p})]d\mathbf{q}\,d\mathbf{p} = \Omega(E)\langle\mathbf{v}\cdot\nabla H\rangle_E,$$

$$\Omega_N(E) \equiv \int(\nabla\cdot\mathbf{v})\,\delta[E-H(\mathbf{q},\mathbf{p})]d\mathbf{q}\,d\mathbf{p} = \Omega(E)\langle\nabla\cdot\mathbf{v}\rangle_E.$$

Now integration by parts yields $d\Omega_D(E)/dE = \Omega_N(E)$ [18]; and

$$\frac{\Omega_D(E_{k+1})}{\hat{\Omega}(E_{k+1})} - \frac{\Omega_D(E_k)}{\hat{\Omega}(E_k)} = \int_{E_k}^{E_{k+1}}\left[\frac{\Omega_N(E)}{\Omega(E)} - (\log\hat{\Omega})'(E)\frac{\Omega_D(E)}{\Omega(E)}\right]\frac{\Omega(E)}{\hat{\Omega}(E)}dE$$

$$= h_k\left[\langle\nabla\cdot\mathbf{v}\rangle_{E_k<E<E_{k+1}} - \hat{\beta}(E)\langle\mathbf{v}\cdot\nabla H\rangle_{E_k<E<E_{k+1}}\right].$$

Thus, if we set $\hat{\beta}(E) = \langle \nabla \cdot \mathbf{v} \rangle_{E_k < E < E_{k+1}} / \langle \mathbf{v} \cdot \nabla H \rangle_{E_k < E < E_{k+1}}$, $\hat{\Omega}(E_k)$ will asymptotically approach the modified density of states $\Omega_D(E_k)$ instead of $\Omega(E_k)$. Accordingly, the converged energy distribution $\rho(E) \propto \Omega(E)/\Omega_D(E) = 1/\langle \mathbf{v} \cdot \nabla H \rangle_E$ is not necessarily flat, and the resulting $\hat{\Omega}(E)$ should be divided by $\langle \mathbf{v} \cdot \nabla H \rangle_E$ in order to recover $\Omega(E)$.

The two alternative mean force formulas mentioned above correspond to $\mathbf{v}^{(\mathbf{p})} \equiv \mathbf{p}$ with $\mathbf{v}^{(\mathbf{p})} \cdot \nabla H = \mathbf{p}^2/m$, and $\mathbf{v}^{(\mathbf{F})} \equiv \mathbf{F}$ with $\mathbf{v}^{(\mathbf{F})} \cdot \nabla H = -\mathbf{F} \cdot \mathbf{F}$, respectively. Thus, the overall energy distributions $\rho^{(\mathbf{p})}(E) \propto 1/\langle \mathbf{p}^2/m \rangle_E$ and $\rho^{(\mathbf{F})}(E) \propto 1/\langle \mathbf{F} \cdot \mathbf{F} \rangle_E$ are not flat.

We can force a flat distribution by using a "normalized" vector field. For an arbitrary $\mathbf{v}$, we construct $\mathbf{v}' = \mathbf{v}/(\mathbf{v} \cdot \nabla H)$, then[18]

$$\hat{\beta}(E) = \langle \nabla \cdot \mathbf{v}' \rangle_{E_k < E < E_{k+1}} = \langle \nabla \cdot [\mathbf{v}/(\mathbf{v} \cdot \nabla H)] \rangle_{E_k < E < E_{k+1}}$$

yields the unbiased $\hat{\Omega}(E_k)$ and a flat distribution, since $\langle \mathbf{v}' \cdot \nabla H \rangle_E = 1$, and $\Omega_D(E) = \Omega(E)$. The above two examples for $\mathbf{v}^{(\mathbf{p})} = \mathbf{p}$ and $\mathbf{v}^{(\mathbf{F})} = \mathbf{F}$, after the normalization, become Eq. (2') and $\hat{\beta}(E) = \langle \nabla \cdot [\mathbf{F}/(\mathbf{F} \cdot \nabla U)] \rangle_{E_k < E < E_{k+1}}$[18], respectively.

We can show Eq. (12) in a similar way. We define the canonical partition function at a fixed volume $V$ as

$$Q(V) = \exp[-\beta F(V)] = \int_{\mathbf{q},\mathbf{p}} \exp[-\beta H_V(\mathbf{q},\mathbf{p})] d\mathbf{q}\, d\mathbf{p},$$

and its estimate

$$\log \hat{Q}(V) = -\beta \hat{F}(V) = \beta \int^V \hat{p}(V') dV'.$$

Then,

$$\frac{Q(V_{k+1})}{\hat{Q}(V_{k+1})} - \frac{Q(V_k)}{\hat{Q}(V_k)} = \int_{V_k}^{V_{k+1}} \frac{d}{dV}\left[\frac{Q(V)}{\hat{Q}(V)}\right] dV = -\beta \int_{V_k}^{V_{k+1}} [F'(V) - \hat{F}'(V)] V^\alpha \frac{Q(V)}{\hat{Q}(V)} \frac{dV}{V^\alpha}$$

$$= -\beta \int_{V_k}^{V_{k+1}} [F'(V) - \hat{F}'(V)] V^\alpha w(V) dV$$

$$= \beta h_k \left\langle [p(V) - \hat{p}(V)] V^\alpha \right\rangle_{V_k < V < V_{k+1}},$$

where $w(V) \equiv \int_{\mathbf{q},\mathbf{p}} \exp[-\beta H_V(\mathbf{q},\mathbf{p})] d\mathbf{q}\, d\mathbf{p} \Big/ [\hat{Q}(V)V^\alpha] = Q(V)/[\hat{Q}(V)V^\alpha]$ is the overall volume distribution, and we have defined $h_k \equiv \int_{V_k}^{V_{k+1}} w(V) dV$. Thus, in order for $\hat{Q}(V_k)$ to converge to $Q(V_k)$, we need $\hat{p} = \langle p(V) V^\alpha \rangle_{V_k < V < V_{k+1}} / \langle V^\alpha \rangle_{V_k < V < V_{k+1}}$, which becomes Eq. (12) by $p(V) = \langle p_{\text{int.}}(\mathbf{q},\mathbf{p}) \rangle_V$.

## Appendix B. Reweighting for the density of potential-energy states

To prove Eq. (4), we first write down the distribution density of the potential energy $h(U)$ in the multicanonical ensemble:

$$h(U) = \int_{-\infty}^{+\infty} \left( \int_{\mathbf{q},\mathbf{p}} \delta[E - U(\mathbf{q}) - K(\mathbf{p})] \delta[U - U(\mathbf{q})] d\mathbf{q}\, d\mathbf{p} \right) [1/\hat{\Omega}(E)] dE.$$

So

$$h(U) = \int_{-\infty}^{+\infty} \left( \int_{\mathbf{q}} [E - U(\mathbf{q})]^{N_f/2-1} \Theta[E - U(\mathbf{q})] \delta[U - U(\mathbf{q})] d\mathbf{q} \right) [1/\hat{\Omega}(E)] dE$$

$$= \int_{-\infty}^{+\infty} \left( \int_{\mathbf{q}} \delta[U - U(\mathbf{q})] d\mathbf{q} \right) (E - U)^{N_f/2-1} \Theta(E - U) / \hat{\Omega}(E) dE$$

$$= g(U) \int_U^{+\infty} (E - U)^{N_f/2-1} / \hat{\Omega}(E) dE,$$

where we have integrated out momenta in the first step, replaced $U(\mathbf{q})$ by $U$ in the second, and used $g(U) = \int_{\mathbf{q}} \delta[U - U(\mathbf{q})] d\mathbf{q}$ in the last. Dividing both sides by the integral yields Eq. (4).

## Appendix C. Structure-Based protein model

The protein model was described in ref. 31,32; we include it here for completeness. The model includes alpha-carbon atoms only, and builds a potential energy based on the geometry of a given protein structure:

$$U = \sum_{\text{bonds}} k_b(b-b_0)^2 + \sum_{\text{angles}} k_\theta(\theta-\theta_0)^2 + \sum_{\substack{\text{dihedrals}\\n=1,3}} k_\varphi^{(n)}\{1-\cos[n(\varphi-\varphi_0)]\}$$

$$+ \sum_{\substack{\text{contacts}\\i<j-3}} \varepsilon\left[5\left(\frac{\sigma_{ij}}{r_{ij}}\right)^{12} - 6\left(\frac{\sigma_{ij}}{r_{ij}}\right)^{10}\right] + \sum_{\substack{\text{not contacts}\\i<j-3}} \varepsilon\left(\frac{C}{r_{ij}}\right)^{12},$$

where $b$ and $b_0$ are the bond lengths of successive atoms along the chain in the given configuration and the reference values, respectively; similarly, $\theta$ and $\theta_0$ are the angles of three successive atoms, $\varphi$ and $\varphi_0$ are the dihedrals of four successive atoms. For non-bonded interactions, we distinguish pairs of contact and non-contact atoms; in the former case, we apply the fourth term with $\sigma_{ij}$ being the distance between $i$ and $j$ in the reference structure, while in the latter case, we apply the last term. We used the following parameters: $k_b = 100$, $k_\theta = 20$, $k_\varphi^{(1)} = 1$, $k_\varphi^{(3)} = 0.5$, and $C = 4.0$ Å.

We used a simpler criterion than the original one[31,36] to define contact atoms: in the reference structure, if any two non-hydrogen atoms from two different residues have a distance $r_{ij} < r_c$, the corresponding alpha-carbon atoms are contacts. Note that Kouza *et al.*[32] used the distances between alpha-carbon, instead of non-hydrogen, atoms to define contacts. The contacts produced by our criterion appeared to be more similar to those yielded by the CSU server[36].

**Figure Captions**

**FIG. 1.** a) The probability distributions for the energy in the microcanonical, canonical, and multicanonical ensembles. b) The ensemble weights in the canonical and microcanonical ensembles.

**FIG. 2.** Multicanonical molecular dynamics on a Lennard-Jones system by the variable-temperature thermostats discussed in Sec. II.C. a-f) The probability distributions for the energy in the system for different methods. These variable-temperature thermostats produce flat energy distributions. Simple replacement of $\beta$ with $\hat{\beta}(E)$ in the canonical thermostat equations, labeled "Unmodified," does not produce flat energy distributions. The parameters are (a) $Q_1 = 300$, (c) $\Delta_{VR} = 0.01$, and (e) $\zeta = 3$. For the Nosé-Hoover chain variable-temperature thermostat in (a) and (b), $M = 5$ and $Q_k = 1$ for $k \geq 2$. Note in (b) the importance of a small time step if the thermostat mass $Q_1$ is too small. In panel (d), the alternative expression $\tilde{\beta}(E) = N_f / \langle 2K \rangle_E$ with $\Delta = 0.3$ produced a non-flat energy distribution, characterized in Appendix A.

**FIG. 3.** (Color online.) Simulations on the villin headpiece by a structure-based model. a) Estimated inverse temperature, $\hat{\beta}(E)$, as a function of energy for different cutoffs, $r_c$. The default inverse temperature $\beta_d$ and reference inverse temperature $\beta_0$ were the same, and they were set to 1.111 for $r_c = 5$ Å, 0.909 for $r_c = 6$ Å, 0.833 for $r_c = 6.5$ Å, 0.667 for $r_c = 7$ Å, and 0.5 for $r_c = 8$ Å. b) The probability distribution for the energy in the system. c) The values of $\hat{\beta}(E)$ (solid, red), $\beta_E^*$ (dashed), and $-\log \hat{\Omega}(E) + \beta_E^* E$ (dotted, blue).

d) The values of $\log g(U)$ (solid, red) and $-\log g(U) + \beta_U^* U$ (dotted, blue). Panels (b)-(d) were obtained from the simulation with $r_c = 8\,\text{Å}$.

FIG. 4. (Color online.) Molecular dynamics with a flat density distribution on a Lennard-Jones system. a) Flat density distributions for three different variable-pressure barostats. b) Values of $\hat{p}(V)$ (thin lines), $p_V^*$ (dashed), and $\hat{G}(V) = \hat{F}(V) + p_V^* V$ (thick lines) for the three barostats. For the Langevin and MC variable-pressure barostats, velocity-rescaling was used as the independent thermostat with $\Delta t_{\text{VR}} = 0.01$.

# References

[1] H. Mori, Prog. Theor. Phys. **34**, 399 (1965); R. Zwanzig, J. Stat. Phys. **9**, 215 (1973).

[2] H. C. Andersen, J. Chem. Phys. **72**, 2384 (1980).

[3] S. Nosé, Mol. Phys. **52**, 255 (1984); C. Braga and K. P. Travisa, J. Chem. Phys. **123**, 134101 (2005).

[4] S. Nosé, J. Chem. Phys. **81**, 511 (1984).

[5] W. G. Hoover, Phys. Rev. A **31**, 1695 (1985).

[6] G. J. Martyna, M. L. Klein, and M. Tuckerman, J. Chem. Phys. **97**, 2635 (1992).

[7] G. J. Martyna, D. J. Tobias, and M. L. Klein, J. Chem. Phys. **101**, 4177 (1994).

[8] D. Frenkel and B. Smit, *Understanding Molecular Simulation From Algorithms to Applications*, Second ed. (Academic Press, 2002).

[9] G. Bussi, D. Donadio, and M. Parrinello, J. Chem. Phys. **126**, 014101 (2007).

[10] B. A. Berg and T. Neuhaus, Phys. Rev. Lett. **68**, 9 (1992); J. Lee, Phys. Rev. Lett. **71**, 211 (1993).

[11] P. M. C. de Oliveira, T. J. P. Penna, and H. J. Herrmann, Eur. Phys. J. B **1**, 205 (1998); J. S. Wang, Eur. Phys. J. B **8**, 287 (1999); J.-S. Wang and R. H. Swendsen, J. Stat. Phys. **106**, 245 (2002); N. Nakajima, H. Nakamura, and A. Kidera, The Journal of Physical Chemistry B **101**, 817 (1997).

[12] F. G. Wang and D. P. Landau, Phys. Rev. Lett. **86**, 2050 (2001); J. Kim, J. E. Straub, and T. Keyes, Phys. Rev. Lett. **97**, 050601 (2006).

[13] J. Kim, J. E. Straub, and T. Keyes, J. Chem. Phys. **126**, 135101 (2007).

[14] Q. Yan and J. J. de Pablo, Phys. Rev. Lett. **90**, 035701 (2003).

[15] V. Martin-Mayor, Phys. Rev. Lett. **98**, 137207 (2007).

[16] L. D. Landau and E. M. Lifshits, *Statistical Physics*, 3rd ed. (Pergamon Press, Oxford, 1980); S.-K. Ma, *Statistical Mechanics*. (World Scientific, Philadelphia, 1985).

[17] D. A. McQuarrie, *Statistical Mechanics*. (Harper & Row, New York, 1976).

[18] H. H. Rugh, Phys. Rev. Lett. **78**, 772 (1997); B. D. Butler, G. Ayton, O. G. Jepps, and D. J. Evans, J. Chem. Phys. **109**, 6519 (1998); O. G. Jepps, G. Ayton, and D. J. Evans, Phys. Rev. E **62**, 4757 (2000).

[19] M. Fasnacht, R. H. Swendsen, and J. M. Rosenberg, Phys. Rev. E **69**, 056704 (2004).

[20] D. J. Earl and M. W. Deem, The Journal of Physical Chemistry B **109**, 6701 (2005).

[21] W. H. Press, S. A. Teukolsky, W. T. Vetterling, and B. P. Flannery, *Numerical Recipes in C: the Art of Scientific Computing*, 2nd ed. (Cambridge University Press, Cambridge, 1992).

[22] Q. Yan, T. S. Jain, and J. J. de Pablo, Phys. Rev. Lett. **92**, 235701 (2004).

[23] C. Zhang and J. Ma, Phys. Rev. E **76**, 036708 (2007).

[24] G. Bussi, T. Zykova-Timan, and M. Parrinello, J. Chem. Phys. **130**, 074101 (2009).

[25] G. J. Martyna, M. E. Tuckerman, D. J. Tobias, and M. L. Klein, Mol. Phys. **87**, 1117 (1996).

| 26 | S. E. Feller, Y. Zhang, R. W. Pastor, and B. R. Brooks, J. Chem. Phys. **103**, 4613 (1995). |
| 27 | M. Chopra and J. J. de Pablo, J. Chem. Phys. **124**, 114102 (2006); N. Rathore, T. A. Knotts, and J. J. de Pablo, Biophys. J. **85**, 3963 (2003). |
| 28 | Y. Duan and P. A. Kollman, Science **282**, 740 (1998); P. L. Freddolino and K. Schulten, Biophys. J. **97**, 2338 (2009); K. Lindorff-Larsen, S. Piana, R. O. Dror, and D. E. Shaw, Science **334**, 517 (2011). |
| 29 | C. Zhang and J. Ma, J. Chem. Phys. **132**, 244101 (2010). |
| 30 | N. Go, Annu. Rev. Biophys. Bioeng. **12**, 183 (1983); M. Kouza and U. H. E. Hansmann, J. Chem. Phys. **134**, 044124 (2011). |
| 31 | C. Clementi, H. Nymeyer, and J. N. Onuchic, J. Mol. Biol. **298**, 937 (2000); H. Lammert, A. Schug, and J. N. Onuchic, Proteins: Struct., Funct., Bioinf. **77**, 881 (2009). |
| 32 | M. Kouza, C.-F. Chang, S. Hayryan, T.-h. Yu, M. S. Li, T.-h. Huang, and C.-K. Hu, Biophys. J. **89**, 3353 (2005). |
| 33 | J. K. Johnson, J. A. Zollweg, and K. E. Gubbins, Mol. Phys. **78**, 591 (1993). |
| 34 | J. Kim and J. E. Straub, J. Chem. Phys. **133**, 154101 (2010). |
| 35 | M.-H. Hao and H. A. Scheraga, J. Phys. Chem. **98**, 4940 (1994); J. Kim, J. E. Straub, and T. Keyes, Phys. Rev. E **76**, 011913 (2007). |
| 36 | V. Sobolev, A. Sorokine, J. Prilusky, E. E. Abola, and M. Edelman, Bioinformatics **15**, 327 (1999). |
| 37 | A. B. Adib and C. Jarzynski, J. Chem. Phys. **122**, 14114 (2005). |

[38] The code for thermostats and barostats is available at

http://www.mwdeem.rice.edu/tbstat

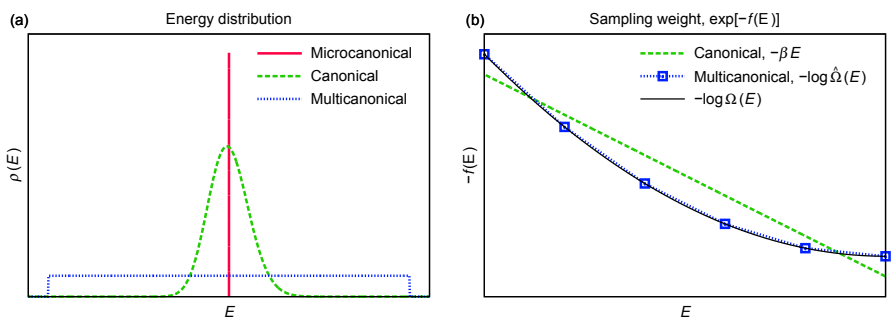

Figure 1:

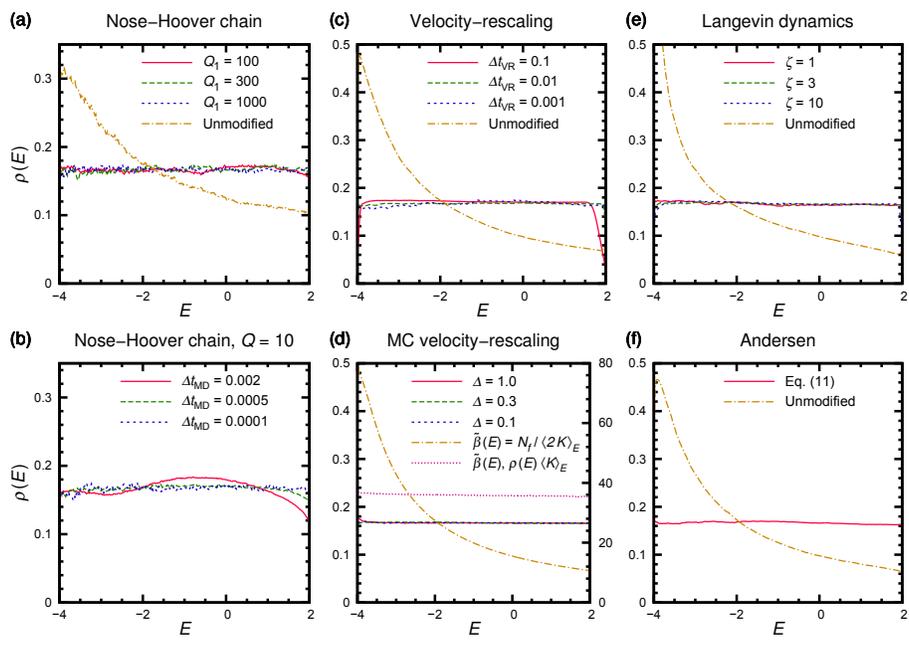

Figure 2:

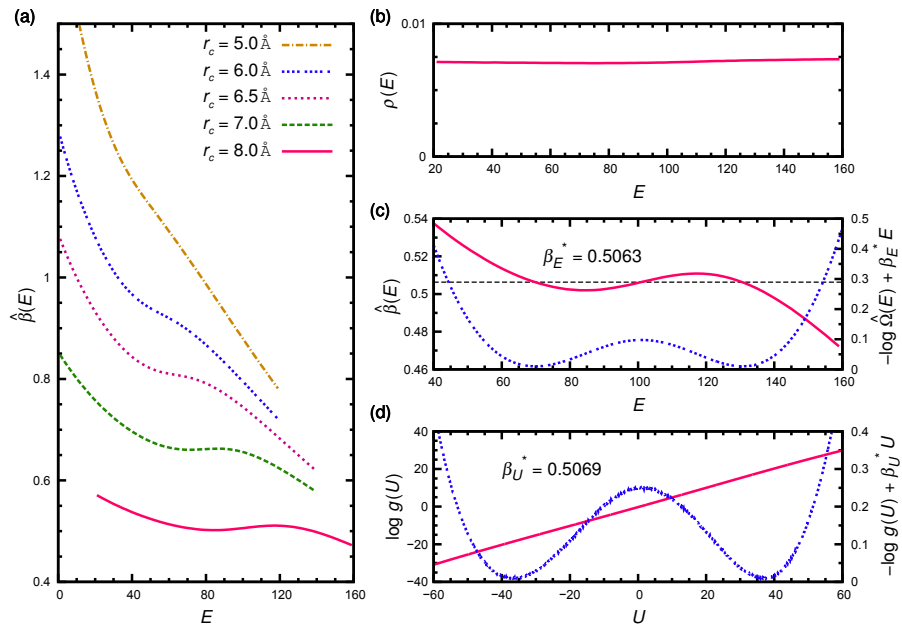

Figure 3:

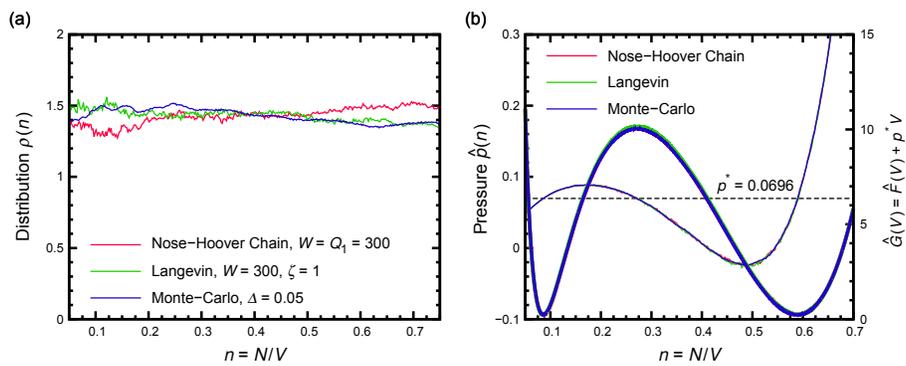

Figure 4: